\documentclass[pre,showpacs,twocolumn,superscriptaddress,epsfig,longeq]{revtex4}

\usepackage{color}
\usepackage{graphicx}
\usepackage{color}
\graphicspath{{./figures/}}

\begin{document}
\title{Dynamic shear jamming in dense granular suspensions under extension}
\author{Sayantan Majumdar}
\email[]{majumdar@uchicago.edu}
\affiliation{James Franck Institute, The University of Chicago, Chicago, Illinois 60637, USA.}
\author{Ivo R. Peters}
\affiliation{James Franck Institute, The University of Chicago, Chicago, Illinois 60637, USA.}
\affiliation{Engineering and the Environment, University of Southampton, Highfield, Southampton S017 1BJ, UK}
\author{Endao Han}
\affiliation{James Franck Institute, The University of Chicago, Chicago, Illinois 60637, USA.}
\author{Heinrich M. Jaeger}
\affiliation{James Franck Institute, The University of Chicago, Chicago, Illinois 60637, USA.}
\date{\today}
\begin{abstract}
Unlike dry granular materials, a dense granular suspension like cornstarch in water can strongly resist extensional flows. At low extension rates, such a suspension behaves like a viscous fluid, but rapid extension results in a response where stresses far exceed the predictions of lubrication hydrodynamics and capillarity. To understand this remarkable mechanical response, we experimentally measure the normal force imparted by a large bulk of the suspension on a plate moving vertically upward at controlled velocity. We observe that, above a velocity threshold, the peak force increases by orders of magnitude. Using fast ultrasound imaging we map out the local velocity profiles inside the suspension, which reveal the formation of a growing jammed region under rapid extension. This region interacts with the rigid boundaries of the container through strong velocity gradients, suggesting a direct connection to the recently proposed shear-jamming mechanism.  
\end{abstract}
\maketitle

\section{Introduction}
Dense suspensions of hard particles in a simple liquid have the remarkable ability to transform from fluid to solid-like behavior when stressed. One well-known trigger for this solidification can be impact on the surface of such fluid \cite{waitukaitis2012impact}, but perhaps even more intriguing is that solidification can also be observed under sufficiently fast extension \cite{white2010extensional, smith2010dilatancy, smith2014fracture}. Recent work has shown that impact-activated solidification (IAS) is a dynamic phenomenon connected to a rapidly propagating jamming front \cite{waitukaitis2012impact, peters2014quasi, peters2016direct}. Here we consider the question whether similar jamming fronts also are responsible for solidification under extension. 
 
As jamming fronts propagate into the bulk of a suspension, starting from the place at the suspension surface where strain was applied, they convert fluid material ahead of them into solid-like material in their wake. If such fronts are indeed created also under extension, then a situation would have to exist that at first glance appears quite counterintuitive: When the suspension surface is pulled upward,  the front would need to move downward rather than propagating in the direction of the applied strain, as for IAS.
 
\begin{figure} 
\noindent\includegraphics[width=8.5cm]{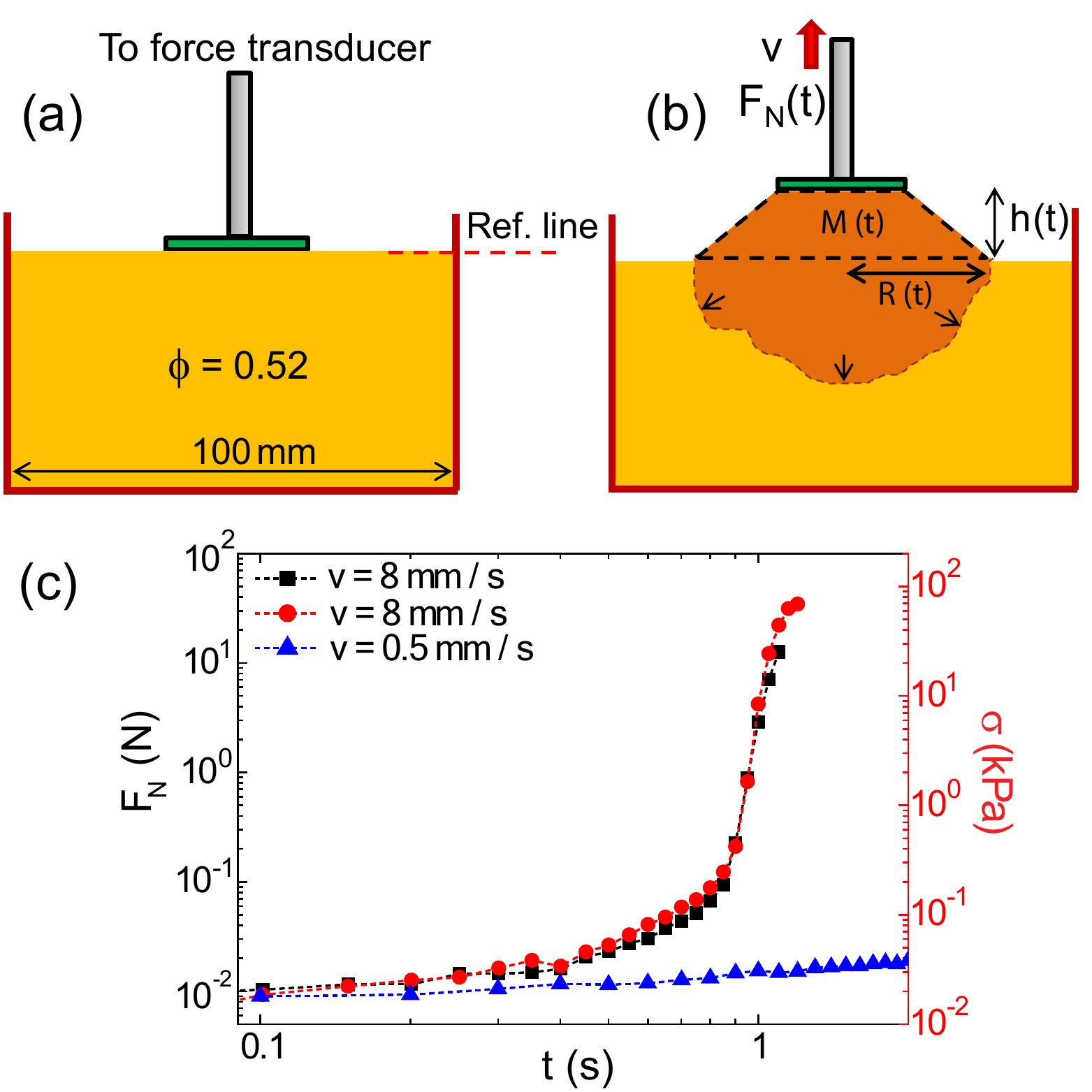}
\caption{\label{F1} (a) Schematic of the experimental set-up. (b) Formation of growing jamming front (brown colored region) under pulling. (c) Normal force and stress as a function of time for two pulling velocities. For $v$ = 8 mm/s, cases of weaker (black squares) and stronger (red circles) adhesion between plate and suspension are shown, $\phi$ = 0.52, $\eta_s$ = 63 mPa\,s.}
\end{figure}
\begin{figure*}
\noindent\includegraphics[width=16.5cm]{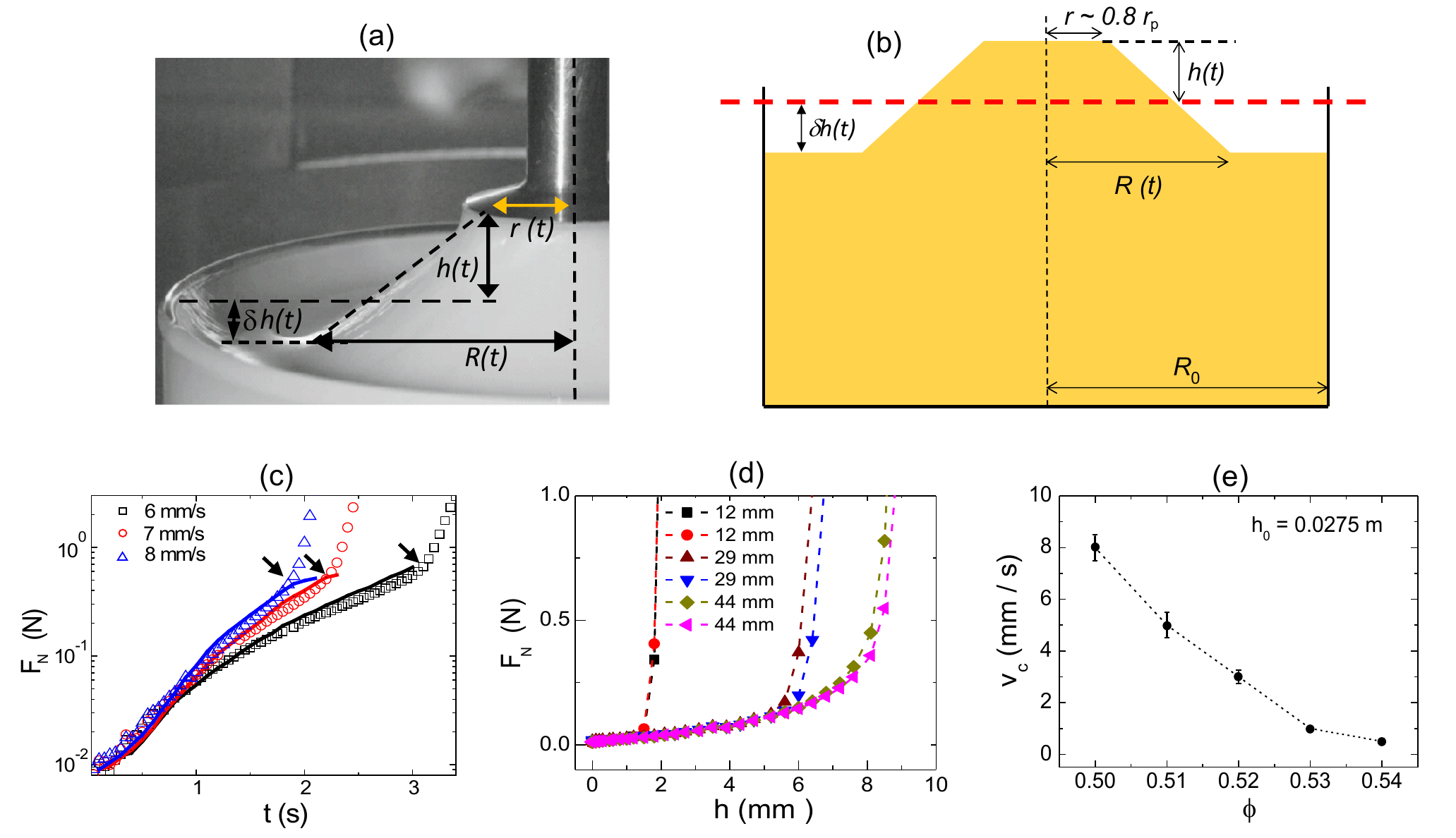}
\caption{\label{F2} (a) Geometrical parameters of the frustum shape formed by the suspension. (b) Schematic representation of the volume pulled out of the bulk and free surface depression under extensile deformation. The frustum shape is approximated by drawing a tangent in the vertical plane on the side surface of jammed region. The instantaneous point of intersection of the tangent at the plate gives $r$ and the intersection point on the free surface of the suspension gives $R(t)$. The red dashed-line gives the initial position of the free surface of the suspension. (c) Force response as a function of time for different pulling velocities. Arrows indicate the onset of rapid increase in force. Solid lines indicate the model predictions. Here, $\phi$ = 0.51 and $\eta_s$ = 50 mPa\,s. (d) Force response as a function of displacement for different initial depths ($h_0$) of the suspension. For each depth, two consecutive measurements are shown. Here, $\phi$ = 0.52 and $v$ = 8 mm/s. (e) Critical plate velocity ($v_c$) for the onset of sharply increasing normal force for different volume fractions $\phi$, $h_0$ = 27.5 mm. The error bars correspond to the minimum velocity step size in going from a viscous like to a sharply increasing force response (each measured three times). The solvent viscosity $\eta_s$ = 20 mPa\,s for both (d) and (e).}
\end{figure*}

\begin{figure}
\noindent\includegraphics[width=8.8cm]{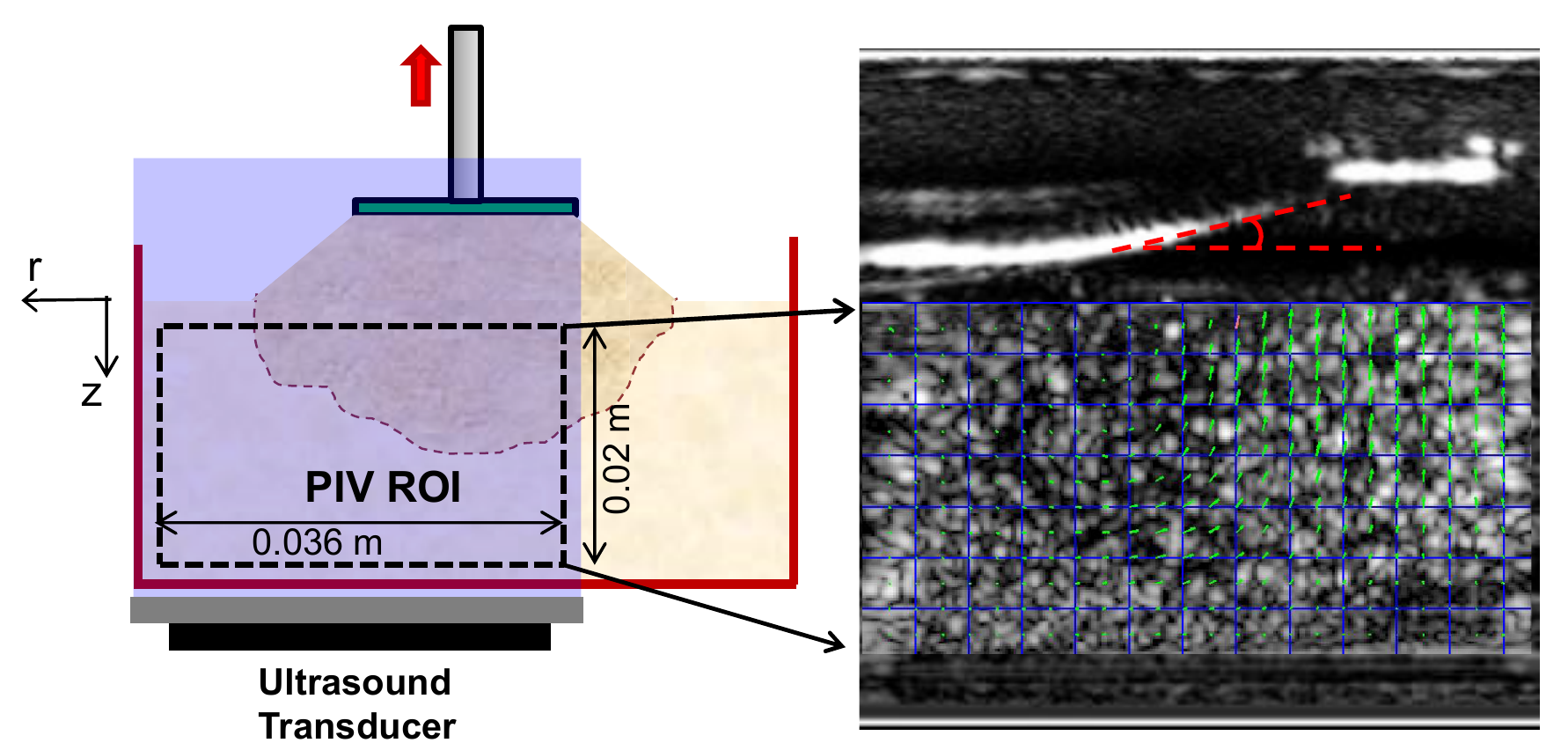}
\caption{\label{F3} Schematics of in-situ ultrasound imaging. The plane of propagation of ultrasound is same as the vertical plane passing through the center of the top plate. A typical ultrasound image with PIV vectors estimated over the Region Of Interest (ROI) is superposed (green arrows) at a particular instant. The angle of the frustum at the free surface is found to be $\sim 16^{\circ}$. The tangent of this angle (or, the slope of the sidewall of the frustum, tan$ 16^{\circ} \sim 0.3$) gives a value very close to the strain required to jam the system. Here, $\phi$ = 0.50 and $\eta_s$ = 10 mPa\,s. The slope of the sidewall of the frustum is found to decrease with increasing packing fraction and pulling velocity.}
\end{figure}

\begin{figure*}
\noindent\includegraphics[width=17 cm]{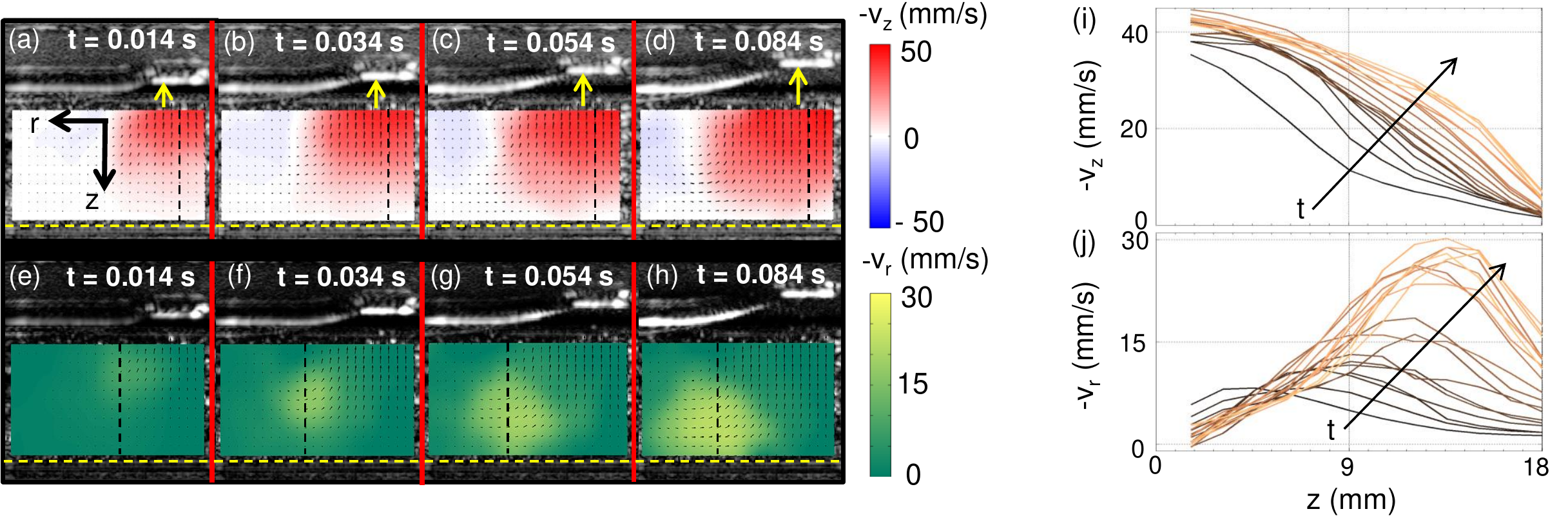}
\caption{\label{F4}  (a) - (d) Time evolution of the flow field under extensional flow. Top plate starts to move at t = 0 s with velocity $v$ = 50 mm/s. Color indicates the vertical components ($-v_z$) of velocity. (e) - (h) Same vector field as before but color indicating the radial ($-v_r$) velocity components. Time evolution of  (i) $-v_z$ and (j) $-v_r$ as a function of z (computed along the vertical dashed line in each panel); Dark to light colors in (i) and (j) indicate increasing time from t = 0.009 s to t = 0.104 s in steps of 0.005 s.  Here, $\phi$ = 0.50 and $\eta_s$ = 10 mPa\,s. }
\end{figure*}

\begin{figure}
\noindent\includegraphics[width=5.75cm]{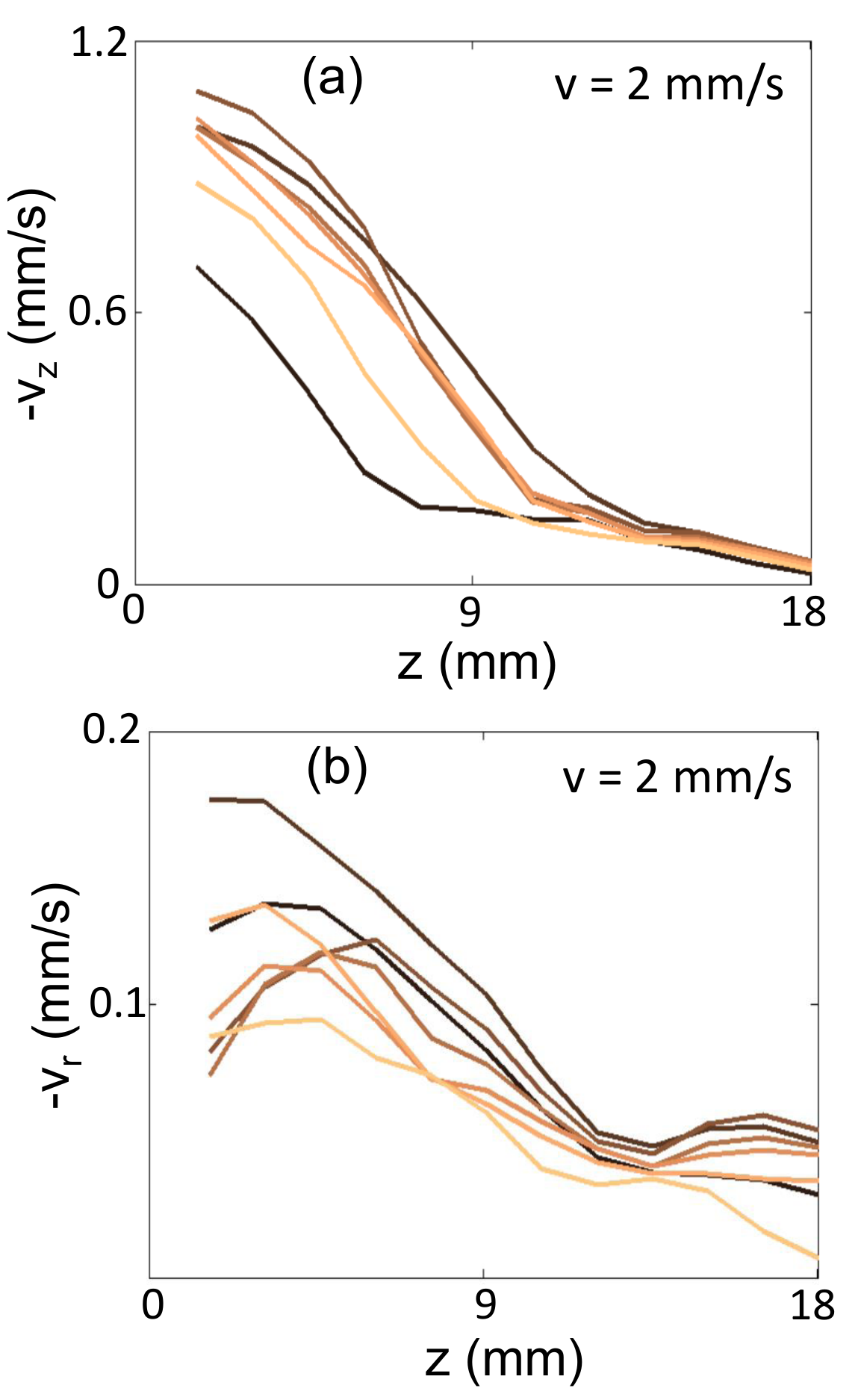}
\caption{\label{F5} (a) Vertical velocity components ($-v_z$) estimated along the vertical line passing through the center of the plate at different instants of time. (b) Radial components of velocity ($-v_r$) along the vertical line that contains the maximum magnitude $|v_r|$ at different instants. For both (a) and (b), the pulling velocity is 2 mm/s and dark to light colors indicate the increasing time from t = 0 s to t = 0.75 s with a time step of 0.125 s. These plots indicate that the high velocity region never reaches the boundary for slow pulling velocities where no rapid force increase is observed. Here, $\phi$ = 0.50 and $\eta_s$ = 10 mPa\,s. Also, both $|\frac{\delta v_z}{\delta z}|$ and $|\frac{\delta v_r}{\delta z}|$ remain $<< 5 s^{-1}$, indicating that the stress scales are well below the jamming transition.}
\end{figure}

\begin{figure}
\noindent\includegraphics[width=7cm]{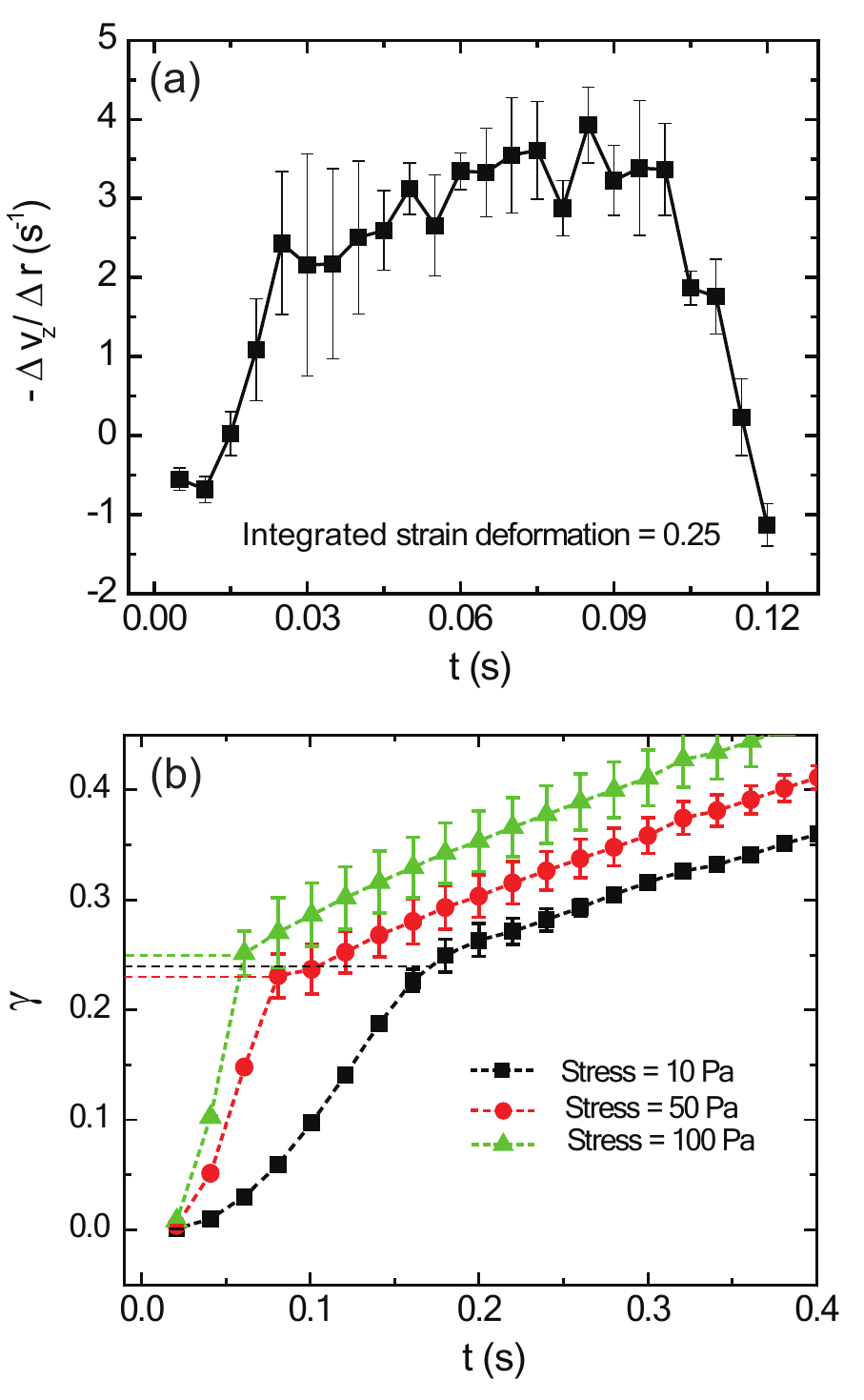}
\caption{\label{F6} (a) Local strain rate (averaged over three consecutive PIV vectors in the vertical direction) as a function of time during jamming front propagation. The parameter values are same as that used in Fig. 4. The error bars are given by the standard deviation of the strain rate at each time estimated at these three windows. The integrated area under the curve gives the total local strain deformation to be $\sim$0.25.
(b) Strain evolution as a function of time for applied stresses (indicated in the figure legend) high enough to jam the system. These experiments are done on the same sample as in part (a) but in a narrow-gap parallel plate rheometer. The onset strain for jamming (indicated by a sudden decrease in slope) is found to be very close to that obtained in part (a). 
}
\end{figure}

\begin{figure}
\noindent\includegraphics[width= 6.5cm]{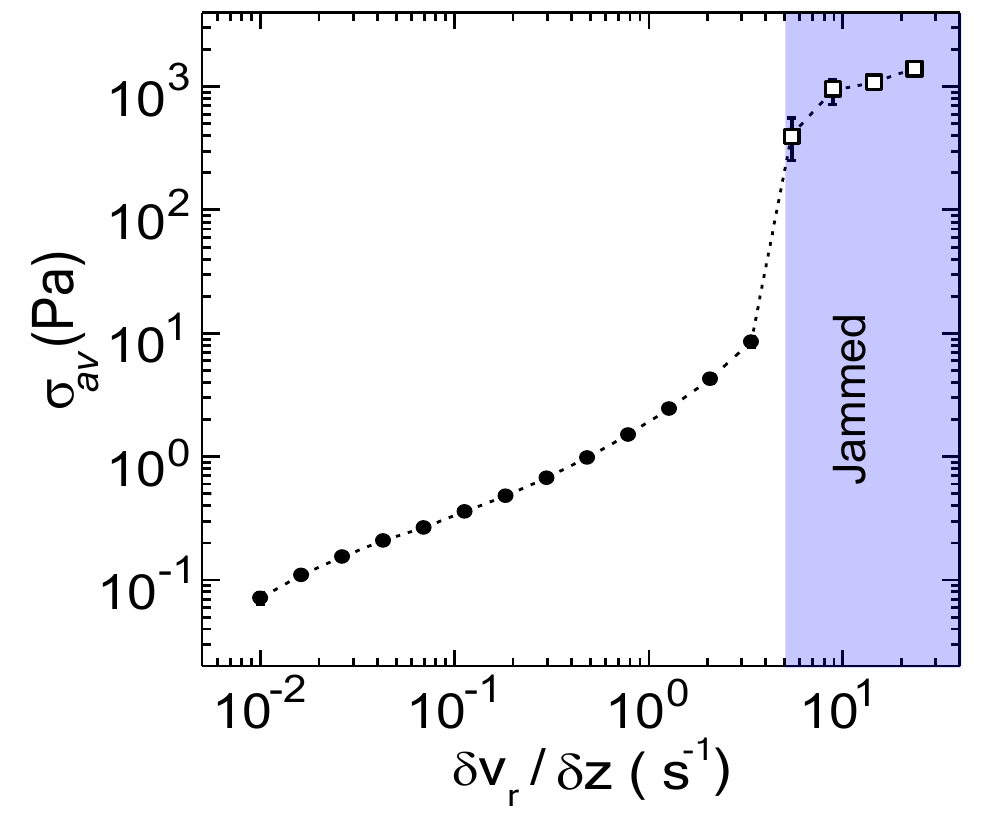}
\caption{\label{F7} Shear-stress vs shear-rate measured under simple shear in a narrow gap rheometer. The shaded region indicates the shear-jammed state. Error bars indicate the standard deviations of three consecutive measurements. Here, $\phi$ = 0.50 and $\eta_s$ = 10 mPa\,s. }
\end{figure}

\begin{figure*}
\noindent\includegraphics[width=12.5cm]{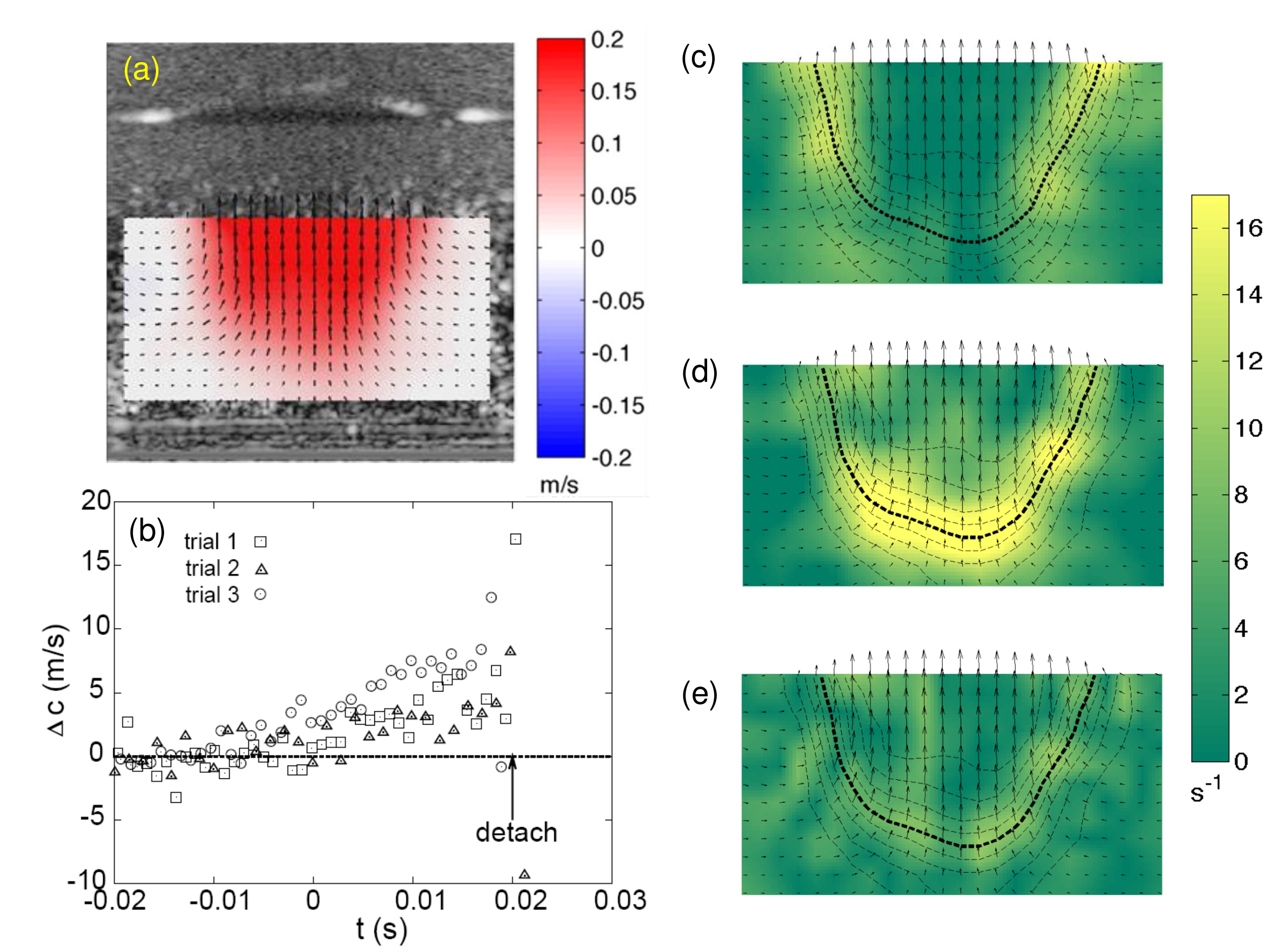}
\caption{\label{F8} Jamming front propagation under extension in a container having a diameter of 10 cm which is $\sim$17 times larger than the diameter of the pulling rod (6 mm). (a) PIV vector field at time t = 10 ms after the top plate starts to move with a velocity $v$ = 0.2 m/s. The vertical components of the velocity are color coded. (b) Change in speed of sound through the suspension during jamming front propagation. The plate starts to move at time t = 0 s with a velocity $v$ = 0.2 m/s and detaches at t $\sim$ 0.02 s. Data for three trials are shown. (c), (d) and (e) Components of strain rate tensor ($\dot{\epsilon}_{r \theta z}$) in cylindrical coordinate system ($\theta$ is the azimuthal angle) for the velocity field shown in (a): (c) magnitude of the simple shear component $|\dot{\epsilon}_{rz}|$, (d) magnitude of the pure shear component $|\dot{\epsilon}_{zz}|$ and (e) magnitude of the rate of expansion component $|\dot{\epsilon}_{rr} + \dot{\epsilon}_{\theta \theta} + \dot{\epsilon}_{zz}|$. The thin dotted lines in (c), (d) and (e) are the contours of points where the vertical velocity drops from 0.9$v$ (uppermost contour) to 0.1$v$ (lowermost contour), in steps of 0.1$v$. The thick dotted line indicates the contour where the velocity drops to 0.5$v$, giving the position of the jamming front. All plots correspond to t = 10 ms after the start of extensional flow. Here, $\phi$ = 0.49 and $\eta_s$ = 5 mPa\,s. }
\end{figure*}

In this paper we investigate the transient dynamics of a large volume of suspension under extensional flow. We report the first observation of propagating jamming fronts under rapid extension. These fronts form the leading edge of a solid-like region that grows into the bulk of the suspension at a rate faster than the extension rate. Mapping out the flow field in the interior of the suspension with high-speed ultrasound imaging, we explicitly show that the fronts coincide with strong, localized shear. This suggests that the concept of shear jamming, originally developed for dry granular materials \cite{cates1998jamming, kumar6167memory, bi2011jamming, vinutha2016disentangling} but recently also identified as relevant for dense suspensions \cite{peters2014quasi, seto2013discontinuous, wyart2014discontinuous, fall2015macroscopic, peters2016direct} applies here as well, although suitably modified to account for the non-steady-state nature of the front propagation process.
 
When a front reaches the nearest rigid boundary of the container, the upward pulling force sharply increases, concomitant with a rapid slowing down of the entire velocity field inside the suspension. The observed peak stress can reach $\approx$ 70 kPa (Fig. \ref{F1}) transiently. Such value is much higher than reported previously for comparatively small cylindrical volumes under extension \cite{white2010extensional}. It is incompatible with explanations based on hydrodynamic lubrication between suspended particles \cite{brown2014shear} and also significantly exceeds capillary stresses due to menisci at the suspension surface \cite{brown2012role}. Here, we show that, such high values of peak stress can be rationalized by a model that treats the jammed region as a fully saturated porous medium and considers the stress required to displace interstitial liquid relative to the grains. 

\begin{figure}
\noindent\includegraphics[width=8.5cm]{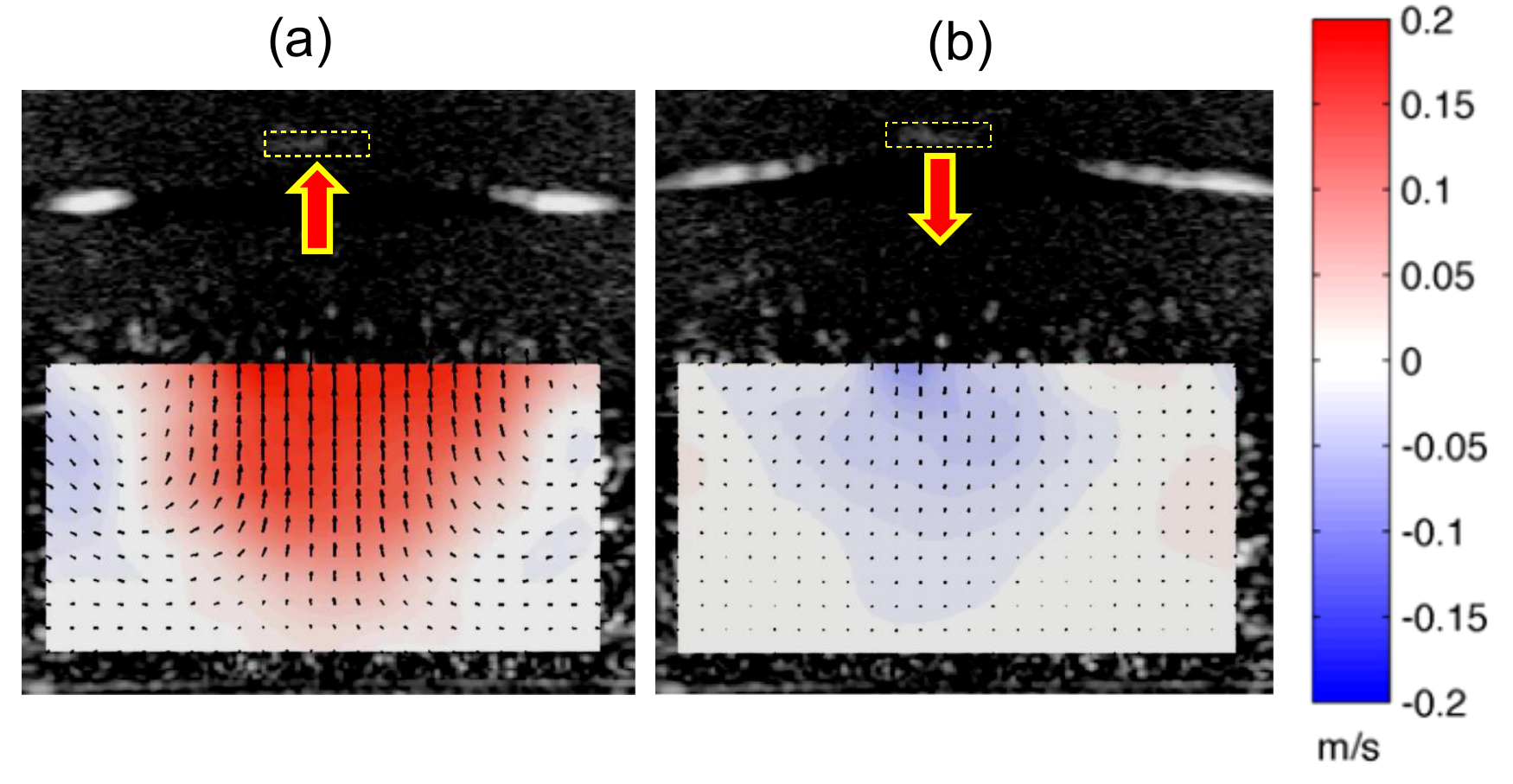}
\caption{\label{F9} Immediate reversal of pulling direction causes the jammed region to disappear. The flow field obtained by ultrasound imaging when the direction of pulling rod (indicated by dashed box) is reversed from upward (a) to downward direction (b). The vertical component of the velocity is color coded. Here, $\phi$ = 0.49 and $\eta_s$ = 5 mPa\,s.}
\end{figure}

\begin{figure}
\noindent\includegraphics[width=6.5cm]{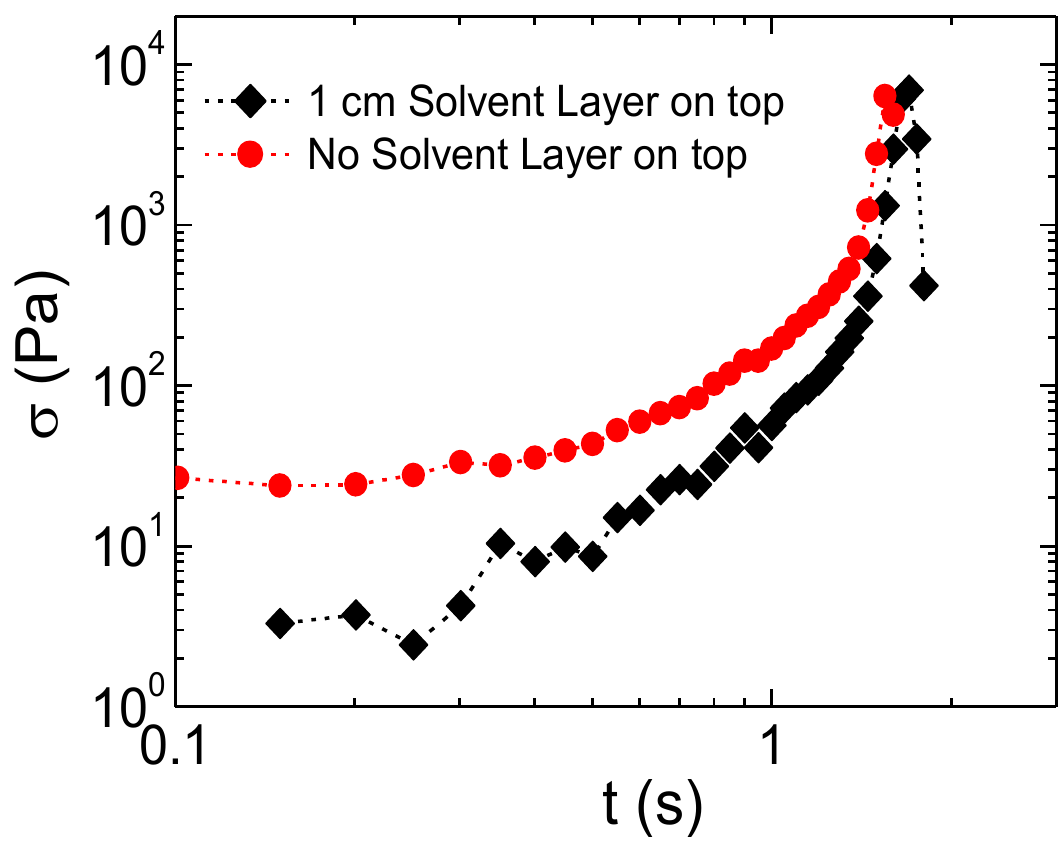}
\caption{\label{F10}  Stress response with (black diamonds) and without (red circles) a solvent layer on top. Here, $\phi$  = 0.51, $\eta_s$ = 50 mPa\,s and $v$ = 8 mm/s.}
\end{figure}

\section{Materials and methods}
Our experiments are performed on a commonly used non-Brownian suspension, cornstarch in water that demonstrate shear thickening. Cornstarch particles have irregular shapes and range in size from 5 $\mu$m to 20 $\mu$m \cite{fall2008shear}. Suspensions are prepared with particle volume fractions ranging from $\phi$ = 0.49 - 0.54 (see Methods in Ref. [6]) by slowly adding cornstarch powder (Ingredion) to a density matched ($\rho = 1.6 \times 10^3~\mathrm{kg/m^3}$) solution of water, cesium chloride (CsCl) and glycerol. By varying the glycerol content, we can tune the solvent viscosity. Here, the solvent viscosity ranges from $\eta$ = 5 mPa\,s to 63 mPa\,s. 

To study the force response under extension, we use a rheometer (MCR 301, Anton Paar). Unless otherwise specified, we use a circular plate of diameter 25 mm connected to the rheometer head through a rod (Fig. \ref{F1}a). For a typical measurement, the plate is first placed on the surface of the suspension and the normal force ($F_N$) is measured on the plate as it is pulled upward at constant velocity $v$ (Fig. \ref{F1}b) controlled through a feedback mechanism. With our rheometer the maximum pulling velocity is limited to 8 mm/s. To achieve higher pulling velocities, instead of a rheometer we use a linear actuator (SCN5, Dyadic Systems) coupled to a plate (diameters: 10 mm and 6 mm) as discussed in the later part of the paper.

To visualize the system dynamics under extension, we use fast ultrasound imaging (Verasonics Vantage 128, at up to 4,000 frames per second) together with the linear-actuator set-up. This allows us to map out the flow profile by tracking the scattering from small air bubbles inside the suspension by using particle imaging velocimetry (PIV). In the ultrasound imaging set-up, the transducer is mounted underneath the suspension, facing upward to image a vertical slice in the same plane as the axis of the pulling plate. The PIV analysis window  has size 36 mm$\times$20 mm (Fig. \ref{F3}) with a spatial resolution of 1.5 mm$\times$1.5 mm. 

\section{Results and discussion}
The variation of measured normal force $F_N$ as a function of time is shown for two different pulling velocities in Fig. \ref{F1}c for $\phi$ = 0.52. For slow pulling velocity ($v$ = 0.5 mm/s), the peak stress is found to reach only $\sim$ 50 Pa. Whereas, for higher pulling velocities ($v$ =  8 mm/s) we see a sharp rise in force, when the peak stress can reach $\approx$ 70 kPa. 

We first focus on the part of the suspension that is pulled out and above the bulk. To track the shape of this part, we use high speed video camera (Phantom V12, Vision Research). Fig. \ref{F2}a shows an illustrative example for $v$ = 8 mm/s. For lower pulling velocities ($ v <$ 3 mm/s in this experiment; data not shown, but see S.I. Movie 2) the system behaves like a highly viscous liquid and gradually flows back into the container. For $v >$ 4 mm/s, the contact line of the suspension with the plate remains pinned and only detaches after the rapid increase in force. This results in the formation of a jammed region whose shape we approximate by a truncated cone or frustum (Fig. \ref{F1}b), which grows with time and maintains its shape under gravity (S.I. Movie 1). Together with the high peak stress mentioned earlier, this is clear signature of a solid-like jammed state, because no fluid can maintain such shape under gravity.

\subsection{Modeling the initial force response}
We consider the instantaneous normal force due to gravity on the pulling plate, coming from the frustum shaped jammed region  having mass $m(t)$. To estimate the instantaneous gravitational force on the jammed region, we need to estimate the volume pulled outside the bulk. For this, we need to first take into account the free surface depression under extension. The dashed red line in Fig.  \ref{F2}b indicates the position of the free surface of the suspension before extension. After the start of extensional flow (at t = 0) there is a jammed region pulled out of the bulk and is well approximated by a frustum shape (Fig. \ref{F2}a). From volume conservation, the free surface outside the frustum will undergo gradual depression ($\delta h (t)$). Here, we neglect the small jammed region formed very close to the side boundaries of the container (shown in Fig. \ref{F2}a) due to high shear resulting from the surface depression. From Fig. \ref{F2}b, using volume conservation, we can write :
\begin{eqnarray}
\frac{1}{3}\pi[h(t) + \delta h(t)][{R(t)}^2 + {r}^2 + R(t)r] = \pi{R_0}^2\delta h(t),
\end{eqnarray}
Solving for $\delta h$ we get,
\begin{eqnarray}
\delta h(t) = \frac{{h\,[R(t)}^2 + {r}^2 + R(t)r]}{3{R_0}^2 - {R(t)}^2 - {r}^2 - R(t)r}.
\end{eqnarray}
Next, we calculate the instantaneous gravitational force on this jammed region having mass $m(t)$ moving upward with a constant velocity $v$. The normal force on the plate is given by Newton's second law: 
\begin{eqnarray}
F_N(t) = d[m(t)v]/dt + m(t)g  = v dm(t)/dt + m(t)g
\label{e1}
\end{eqnarray}
Here, the mass $m(t) = V(t)\rho$, where $V(t)$ is the volume of the jammed region pulled out of the bulk and $\rho$ is the density of the suspension and $g$ is the acceleration due to gravity. $V(t)$ can be estimated from geometry: 
\begin{eqnarray}
V(t)\approx\frac{\pi [h(t) + \delta h(t)]}{3}[r(t)^2 + R(t)^2 + r(t)R(t)], 
\label{e2}
\end{eqnarray}
where $r (t)$ and $R(t)$ are the top and bottom radii of the frustum as determined from the tangent drawn at the middle of the curved surface of the frustum in the vertical plane, $h(t)$ is the height of the plate measured from the rheometer data, and $\delta h(t)$ gives the surface depression [Eq.(2)]. To good approximation, the top radius is independent of time, $r(t)\approx 0.8r_p$ ($r_p$: top plate radius, see Fig. 2a). For the range of pulling velocities we use, the term $m(t)g >> v \cdot d[m(t)]/dt $, thus,  $F_N(t)\sim m(t)g$.

We compare the force calculated from this model with the measured force in Fig. \ref{F2}c. We find very good agreement at early times, demonstrating that the initial force evolution is dominated by the growing mass of the jammed region pulled outside the bulk of the suspension.

\subsection{Jamming front propagation}
To explain the sharp rise in force under extension, we assume that, similar to the case of impact, the jamming front grows in all directions, including downward into the bulk (Fig. \ref{F1}b), although the part that is growing inside the bulk cannot be seen optically because of the opaque nature of the suspension. As with IAS, the rapid rise in force then corresponds to the jamming front reaching the container bottom (or a side wall) \cite{waitukaitis2012impact, peters2014quasi}. 

To test these assumptions, we map out the force response by varying the initial depth $h_0$ of the suspension (Fig. \ref{F2}d) for $v$ = 8 mm/s. The critical plate displacement ($h_c$) at which the force starts to rise sharply, increases as we increase $h_0$. Such delayed force rise with increasing suspension depth is consistent with a jamming front that propagates roughly at a constant velocity, similar to the case of impact \cite{waitukaitis2012impact, peters2014quasi, waitukaitis2014impact}. Measuring the position of the front from the moving plate, the ratio of front to plate velocities is given by, $k = \frac{v_f}{v} = 1 + h_0/h_c$. For different $h_0$ values, we find an average $k = 6.5 \pm 0.7$ ($\phi$ = 0.52), indicating that the front travels much faster than the pulling plate as also been observed under impact \cite{waitukaitis2012impact, peters2014quasi, waitukaitis2014impact}. For a given $h_0$, the critical velocity ($v_c$) decreases with increasing volume fraction (Fig. \ref{F2}e).

To understand the correlation between the observed force response and system dynamics, to visualize the jamming fronts, and to further explore connections to jamming  under impact, we use ultrasound imaging. In Fig. \ref{F4}a-d we show the velocity fields at different instants of time inside a suspension with $\phi$ = 0.50 and $\eta_s$ = 10 mPa\,s for a pulling velocity $v$ = 50 mm/s. The z-components of the velocity ($-v_z$) are color coded. We see a correlated region of high velocity that grows with time and finally interacts with the container bottom (horizontal dashed lines in Fig. \ref{F4}a-h) similar to the jammed region observed under impact \cite{peters2014quasi, han2016high}. In Fig. \ref{F4}e-h, we show the same velocity profiles as in Fig. \ref{F4}a-d but now with color indicating the radial velocity ($-v_r$) component. We find that the maximum of $|v_r|$ is localized near the edge of the jammed region and, as this region grows with time, the magnitude of $v_r$ increases. We map out the time evolution of vertical ($-v_z$) and radial ($-v_r$) velocity profiles as a function of depth ($z$) in Fig. \ref{F4}i and Fig. \ref{F4}j, respectively, up to the point beyond which the entire velocity field within the PIV window suddenly slows down. At early times, when the front just starts to grow, the slope (gradient) of the velocity close to the bottom boundary of the container (larger values of $z$) remains negligible compared to that near the plate (smaller values of $z$), but at later times the gradient near the container boundary gradually becomes stronger. The value of this gradient is found to be $\sim 4-5$ s$^{-1}$. In contrast, for slow pulling velocities, where the normal force does not show any rise, the velocity gradient close to the rigid boundaries of the container remains negligible compared to that close to the moving plate at all times (Fig. \ref{F5}).

\begin{figure}
\noindent\includegraphics[width=7cm]{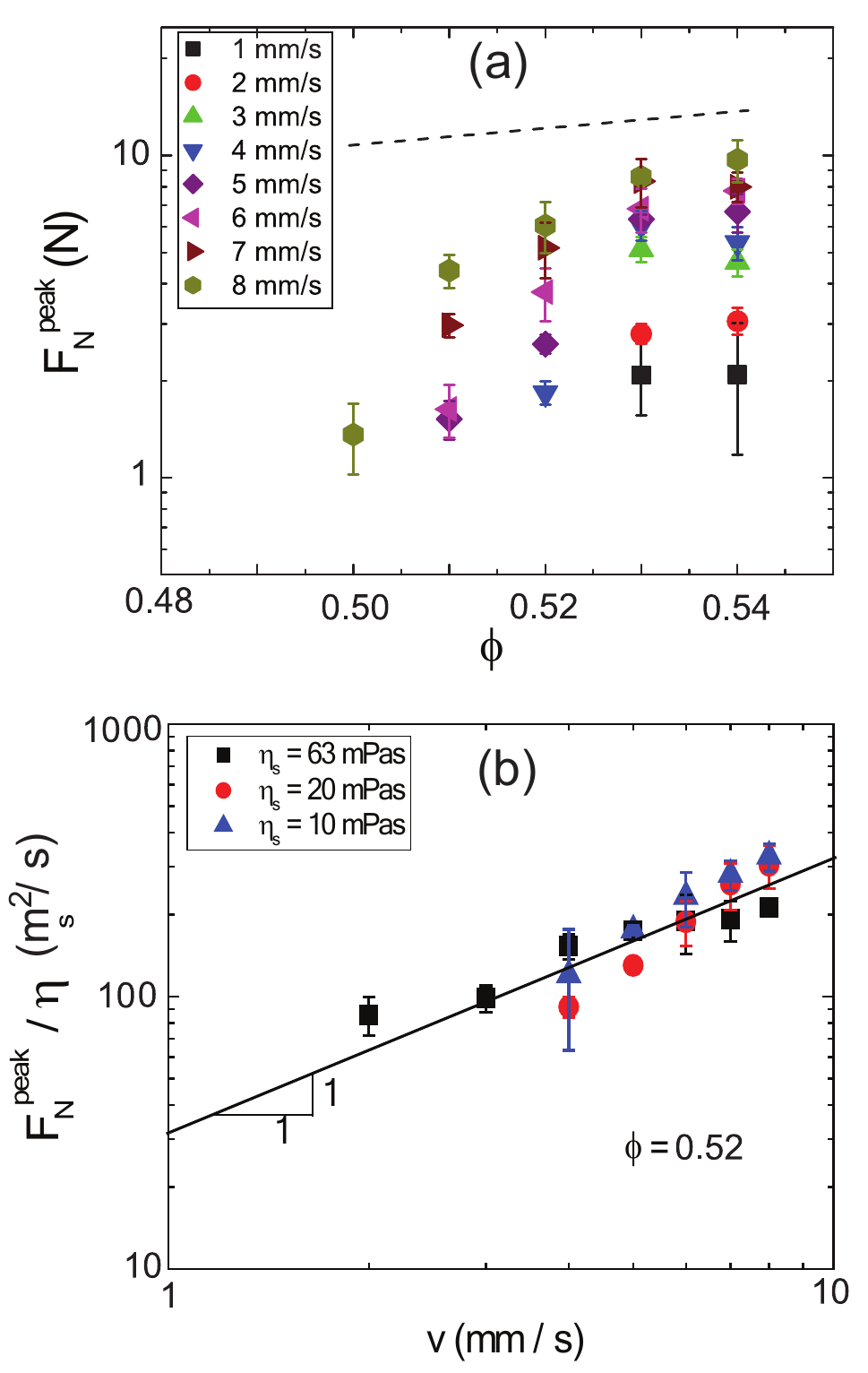}
\caption{\label{F11} (a) Peak force as a function of volume fraction in weak adhesion limit (plate just touches the surface) for different pulling velocities as indicated. The dashed line indicates a function $\propto\, \frac{{\phi}^2}{(1 - {\phi})^3}$; the trend is consistent with K-C relation. Although, an overall increasing trend of peak force is observed as a function of $\phi$, due to the large error bars and small range of $\phi$ (0.50$\,\leq \phi\leq$ 0.54) probed experimentally, the exact functional form is difficult to estimate accurately. Nevertheless, for higher values of pulling velocities and $\phi$, the K-C prediction seems to match the trend in experimental data. (b) Peak stress normalized by solvent viscosity as a function of pulling velocity. The data collapse indicate that the peak stress increases linearly with solvent viscosity and pulling velocity. }
\end{figure}

\begin{figure}
\noindent\includegraphics[width=7cm]{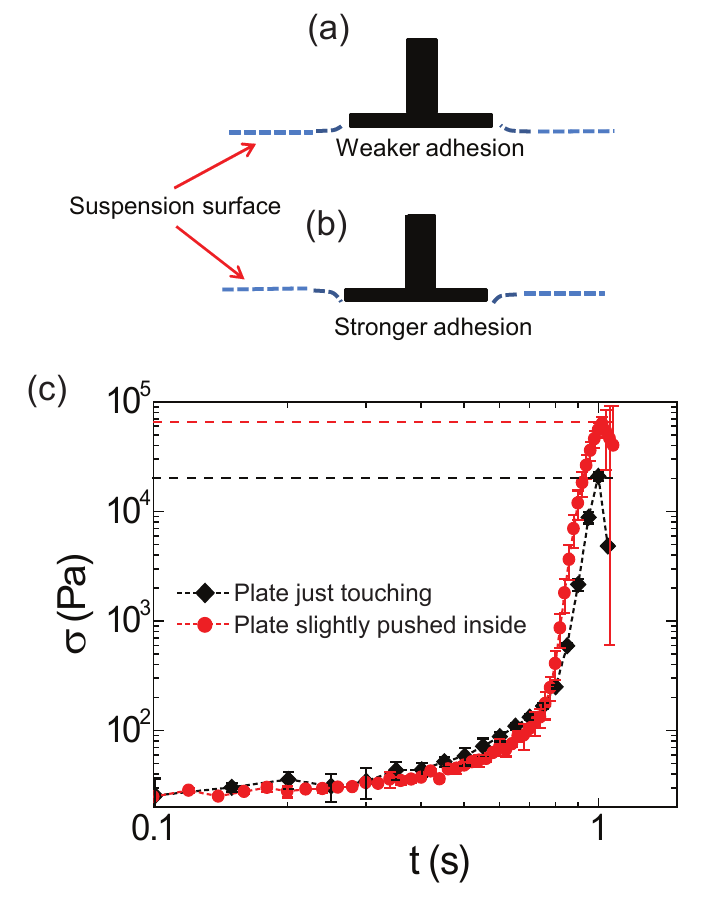}
\caption{\label{F12} Stress response under extension for two slightly different initial positions of the plate: (a) The plate just touches the suspension surface and (b) the plate is slightly ($\approx$2 mm) pushed inside the suspension. (c) The peak stress under extension ($v$ = 8 mm/s) for initial conditions given in (a) (solid black diamonds) and (b) (solid red circles). The measured peak stress is more than three times larger when the plate is pushed inside indicating that the peak stress is governed by the coupling between the plate and the suspension. Here, $\phi$  = 0.54 and $\eta_s$ = 20 mPa\,s. The error bars indicate the average standard deviation from three consecutive measurements.}
\end{figure}

\subsection{Strain field and shear jamming}
Integrating the local shear rate over time, we find (Fig. \ref{F6}a) that the total accumulated strain at an element (corresponding to a single PIV vector) when the jamming front passes through it is $\sim$ 0.25, similar to that found in dry granular systems \cite{bi2011jamming}. We also find that similar strain values are required to jam the system from a conventional start up measurement (Fig. \ref{F6}b). Furthermore, we can obtain it from the slope of the sidewall of the frustum (Fig. \ref{F3}). Thus, in our experiments, the frustum shaped region provides a direct visual estimation of the accumulated strain inside the jammed portion of the suspension. 

The large shear rate at the front implies that the suspension can generate significant stresses, large enough to drive it into a solid-like jammed state. We can see this explicitly from a separate experiment, where the same suspension is subjected to a steady-state shear ramp in a parallel plate rheometer (Fig. \ref{F7}). At shear rates of $\sim 4-5$ s$^{-1}$, corresponding to those at the front, the suspension behavior changes abruptly and the suspension enters a state where homogeneous flow is no longer possible and steady shearing initiates solid-like failure modes, such as partial detachment of the sample from the plate, slippage and formation of cracks (the maximum stress achievable with a transiently shear jammed state cannot be probed with steady state driving because of these failure modes; the measured values [open symbols in Fig. \ref{F7}] thus underestimate this stress). 

Mapping out the full 3-D strain rate tensor \cite{han2016high} we find that a high shear rate (both pure and simple shear) region having much larger magnitude than the rate of expansion component is formed at the edge of the growing jammed region (Fig. \ref{F8} and Movie 3). This boundary shear provides the confining stress for the jammed region. Sound velocity measurements during the jamming front propagation under extension further confirm that there is no significant local change in packing fraction (Fig. \ref{F8}). Up to the point where the frustum detaches, the measured change in sound speed $\Delta c$ is less than 15-20m/s. From Ref. [15] the speed of sound $c$ in similar suspensions is around 1,500 m/s and the relationship between $c$ and $\phi$ is linear. This implies that any packing fraction change could at most have been $\Delta \phi \sim$ 0.01-0.02.  Given that we start in our experiments from a fluid state that has a $\phi$ significantly below the packing fraction for jamming by compression of the particle subphase, such small $\Delta \phi$ is insufficient to solidify the system and instead the jamming is primarily shear induced \cite{han2016high}.

\subsection{Shear jamming in suspensions: general discussion}

The similarity between force response of dense suspension under impact \cite{waitukaitis2012impact, peters2014quasi, waitukaitis2014impact} and extension sheds light on the jamming mechanism. Particularly, it strongly supports the notion of jamming induced by shear \cite{kumar6167memory, bi2011jamming}. In our case,  shear enters into the problem in a non-trivial way: localized strong shear at the boundary between the moving and the (still) unperturbed portion of the suspension gives rise to a growing jammed region through shear-induced network formation between particles \cite{bi2011jamming, seto2013discontinuous, fall2015macroscopic}. The particles experience sufficiently high local stresses to be pushed into frictional contacts, thereby breaching the lubrication barrier between them \cite{wyart2014discontinuous, cates2014granulation}. For the same cornstarch suspension system, jamming by simple shear (without any compression / extension) was demonstrated with a wide gap Couette geometry recently \cite{peters2016direct}, where the breakage of lubrication and onset of jamming are discussed in detail with a steady state phase diagram.

\begin{figure}
\noindent\includegraphics[width=8.5cm]{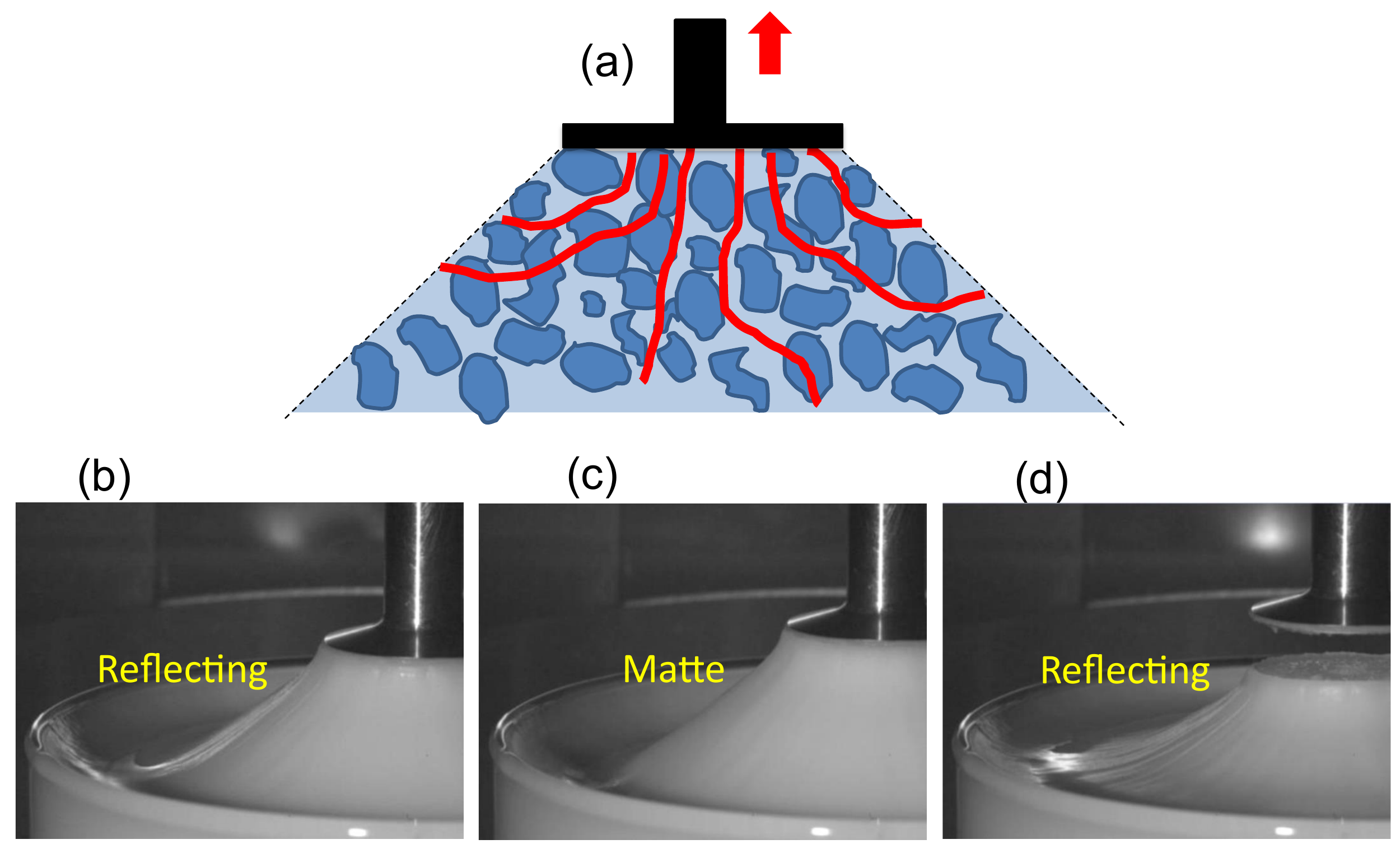}
\caption{\label{F13} (a) Schematic showing the flow of solvent through the jammed granular pack of particles, red lines indicate typical capillary paths consistent with the Kozeny-Carman scheme. The surface texture: Before force shoot up (b), at the onset of force shoot up (c), and after the detachment (d).}
\end{figure}

\begin{figure}
\noindent\includegraphics[width=5.5cm]{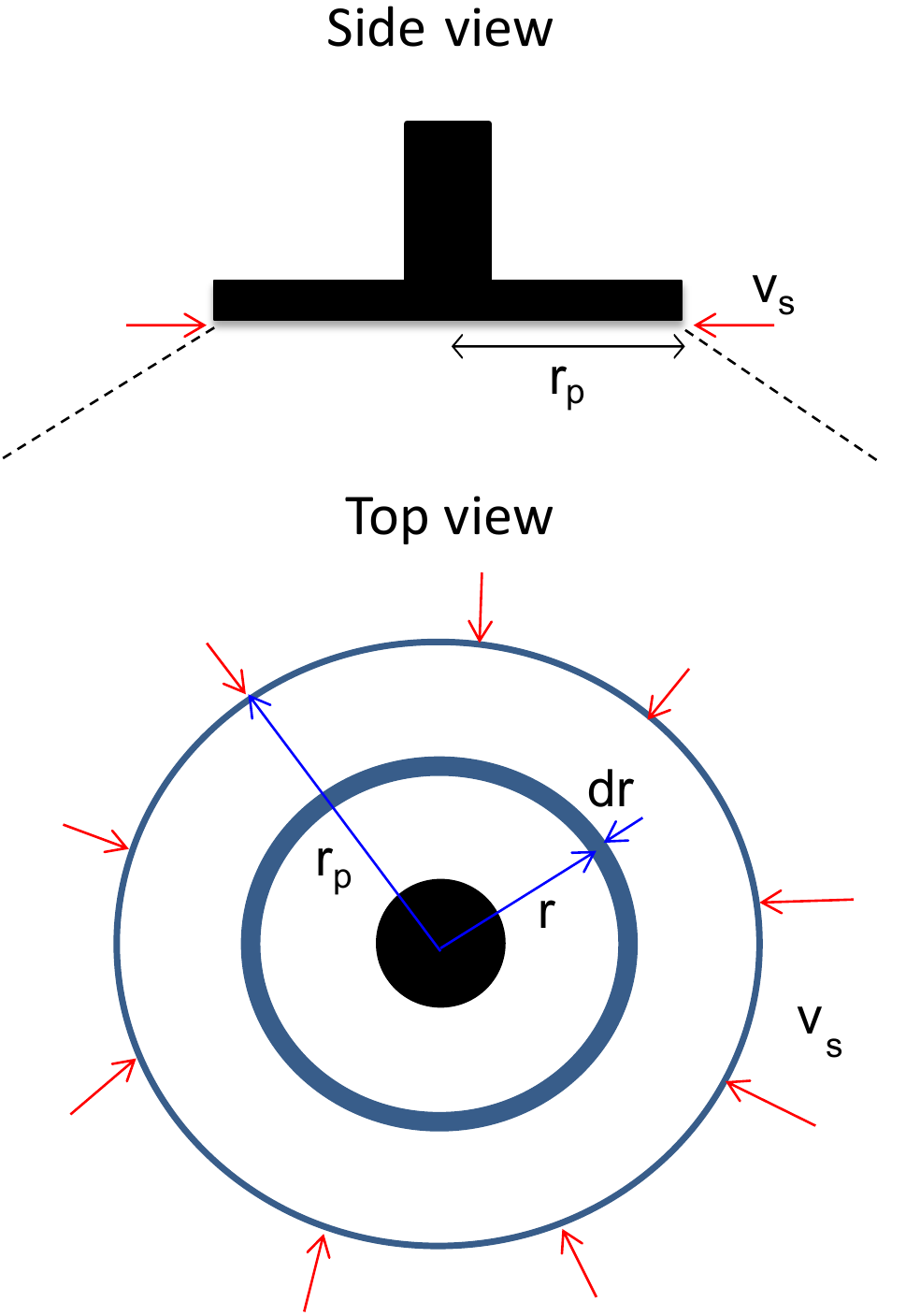}
\caption{\label{F14} Schematics show drainage through the sidewall of the frustum, close to the plate.}
\end{figure}

\begin{figure}
\noindent\includegraphics[width=7cm]{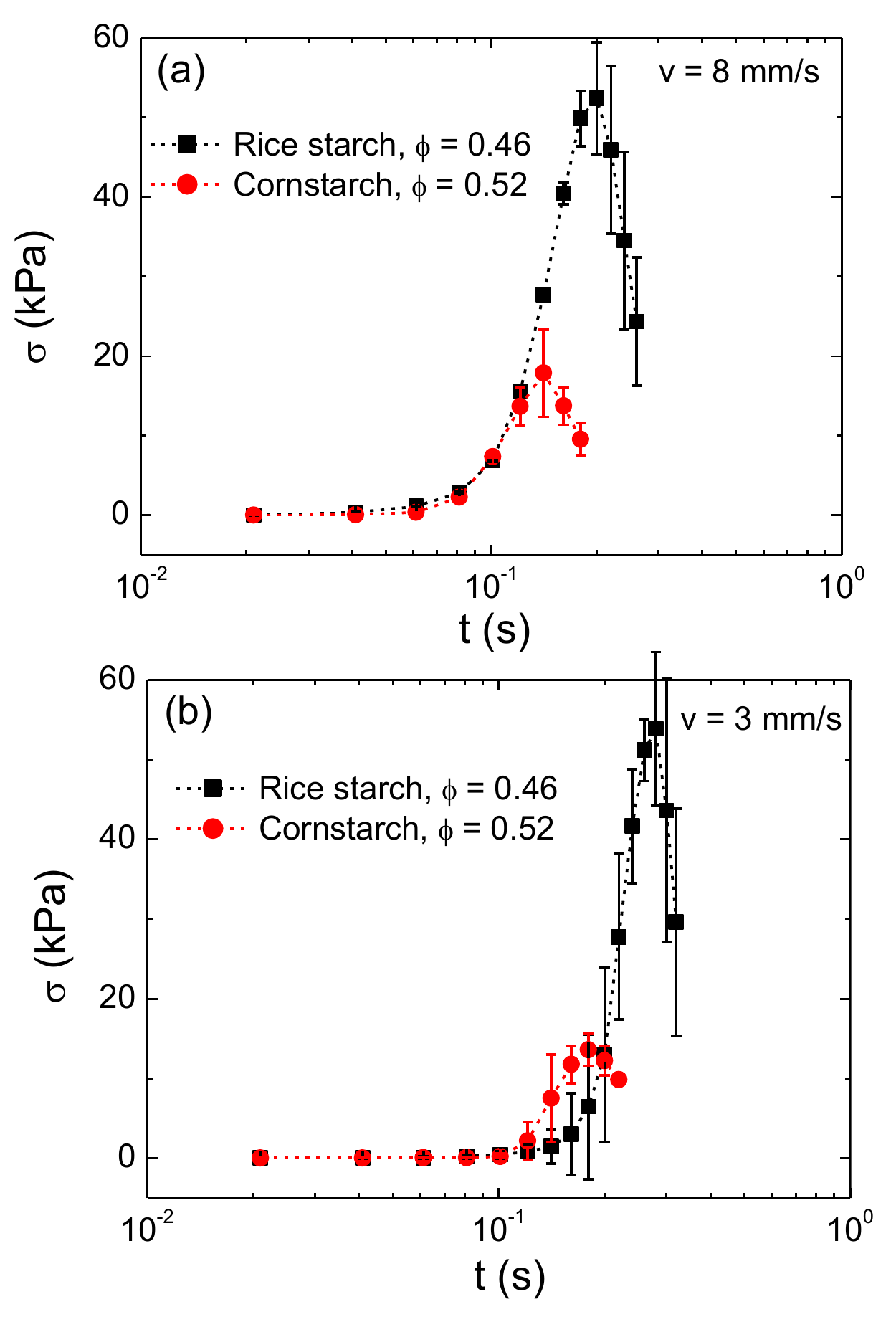}
\caption{\label{F15} Peak stress as a function of time for cornstarch and rice-starch particles. The volume fractions are indicated in the figures for a pulling velocity (a) v = 8 mm/s and (b) v = 3 mm/s . The solvent viscosity $\eta_s$ = 8 mPa\,s in both the cases. The error bars indicate the standard deviation of three independent measurements on the same sample.}
\end{figure}

We find that immediate reversal of the direction of motion of the pulling head causes the jammed region to disappear (Fig. \ref{F9}). Such asymmetric response similar to that observed in \cite{gadala1980shear, lin2015hydrodynamic} under simple shear, indicates the importance of frictional contacts in the observed jamming dynamics. Here, we have used the term `frictional' in a generalized sense that encompasses all short range \emph{contact} interactions between particles. Short range repulsive interactions need to be overcome before the frictional contacts can proliferate. These repulsive interactions set the stress onset for shear thickening / jamming in dense suspensions. However, the exact nature of such inter-particle repulsive interactions can be complicated as they can arise from a variety of sources such as: Brownian motion, zeta potential, steric repulsion etc. \cite{brown2014shear}.  

The stress required for jamming is provided by the inertia of the unperturbed suspension ahead of the front. Inside the bulk of the suspension, therefore, the flow field is essentially the same as under impact, but with the direction of flow reversed. When the jammed region reaches the container boundary, strong squeeze flow generates strong shear that slows down the entire flow field due to shear jamming and gives rise to a solid-like contact network of particles extending from the bottom of the container to the top plate. 

\subsection{Peak stress under extension}
In impact experiments, the rod (impactor) always remains strongly attached to the suspension and the coupling is only limited by the formation of cracks at very high stresses \cite{waitukaitis2014impact}. However, for the case of extension, there is a sudden detachment of the jammed suspension from the pulling plate. For suspensions showing discontinuous shear thickening (DST), the maximum stress scale is provided by the confining capillary pressure at the suspension-air interface \cite{brown2014shear} which is proportional to $\gamma / a$, where $\gamma$ is the surface tension of suspension-air interface and $a$ is the typical diameter of the particles. To test for the role of capillary effects in setting the peak force in our system, we get rid of the effective surface tension of the suspension air interface by pouring a layer ($\sim$ 1 cm deep) of solvent above the suspension so that the plate remains submerged. Force measurement during extension under this condition reveals that the magnitude of peak stress remains essentially the same (Fig. \ref{F10}), indicating that the capillary stress is not important for the observed force response. For our experiments, the observed peak stresses are much higher than the predicted capillary stresses ($< 1$ kPa) \cite{brown2012role}.
Furthermore, the peak normal force is also found to increase linearly with increasing pulling velocity (Fig. \ref{F11}). Thus, under extension the peak stress is set by dynamic effects rather than static capillarity.

In our experiments, the detachment after jamming always happens only at the contact region of the frustum with the top plate. This indicates that the adhesion strength between the plate and the suspension plays an important role in setting the peak stress.
To see the importance of adhesion strength between the plate and the suspension, we compare the average stress response when the the initial position of the plate just touches the suspension (Fig. \ref{F12}, panel a) and when initially the plate is slightly pushed into the suspension (by $\approx 2$mm) (Fig. \ref{F12}, panel b). In both the cases, the contact area between the plate and the suspensions remains effectively the same. We see that (panel c of Fig. \ref{F12}) for $v$ = 8 mm/s ($\phi$ = 0.52; $\eta_s$ = 20 mPa\,s), the peak stress (obtained by averaging over three consecutive runs) is more than three times higher when the plate is pushed slightly inside, indicating that the coupling between the plate and the suspension plays an important role. It should be clear that the adhesion effects come just before the point of detachment and not from the jamming dynamics. This is supported by the fact that within experimental error the two curves match up well before the detachment (Fig. \ref{F12}c). Similar behavior can also be seen in Fig. \ref{F1}c. 

Although the jamming mechanism is similar for both impact and extension, it may seem conceptually difficult to understand how extensional deformation can impart high normal stress on the plate. In the case of impact activated solidification (IAS), once the jamming front hits the container boundary, there is a solid-like jammed region that extends between the point of impact and the container wall. To push the impactor further involves deforming this solid-like jammed region that requires huge amount of stress. This explains the sharp rise in stresses observed in IAS. However, in case of extensile deformations such picture is no longer valid, since we are not pushing against such solid-like jammed region, but rather pulling away from it. To resolve this issue we proposed a dynamic model based on Darcy flow arguments as discussed next.

We note that, just before the detachment, the plate still moves at a constant velocity and for continuity it tends to draw the solvent into the jammed network of particles. This is easily observed by a sudden change in surface texture of the frustum (from shiny to matte) just before the detachment, especially in the region near the plate (Fig. \ref{F13}). 

We estimate the average pressure $\Delta P$ (or stress $\sigma$) at the top plate to suck the solvent (viscosity $\eta_s$) through this jammed, porous medium, using a modified Kozeny - Carman relationship \cite{mccabe2005adsorption, von2011nonmonotonic}
\begin{eqnarray}
\frac{\Delta P}{r_p} = \frac{60 \eta_s {\phi}^2 v_s}{{\Xi}^2 a^2 (1 - {\phi})^3}, 
\label{e3}
\end{eqnarray}
(see Appendix A and Fig. \ref{F14}) where the sphericity parameter $\Xi$ is of order unity, $v_s$ is the solvent velocity, approximated by the plate velocity $v$, and $a$ the particle diameter. We get a typical estimate for stress on the top plate $\sigma\approx$ 3 MPa (see Appendix A and B). However, this theoretical upper limit is unlikely to be reached in practice because of other physical effects. In particular, the lower stress along the rim, due to the reduced path length, leads to breakage of the solvent layer at the rim before the maximum stress is reached. This can give rise to a dynamic delamination leading to eventual full detachment. While the exact mechanism remains to be explored, we believe that the competition between the dynamic delamination process and steady state solvent drainage sets the value of peak stress observed in our experiments. This is corroborated by the observation that the peak stresses scale linearly with $v$ and $\eta_s$ (Fig. \ref{F11}).
\section{Conclusion and outlook}
Our results, and in particular the observation of frustums with clearly defined, angled side walls, directly indicate that sufficiently rapid extension can lead to a transformation of a dense suspension from fluid to solid. This goes beyond shear-thickening and implies dynamic jamming, i.e., the reconfiguration into a system with a finite yield stress.
Our study finds that the transient jamming dynamics under extension in a dense granular suspension in many ways closely resemble those under impact. In particular, in both situations dynamic shear jamming fronts are generated that propagate downward into the bulk of the suspension at speeds larger than the pulling or impact speed. At high pulling speed, the jammed region strongly couples the moving plate and the suspension giving rise to strong shear flow along the propagating front. The extent to which such shear can induce jamming depends, according to current models, on the degree of frictional interaction between particles \cite{seto2013discontinuous, wyart2014discontinuous}. Cornstarch suspensions are among the systems exhibiting the most pronounced shear thickening and shear jamming behavior, which indicates that particle-particle interactions are likely to be highly frictional. However, we expect that similar behavior to that reported here should be observable in other types of suspensions too. Certainly different types of starch show this behavior (Fig. \ref{F15}).

One very interesting question for future studies will be whether the dynamic jamming behavior observed here for non-Brownian suspensions will extend into the regime of dense colloidal systems, where the particles are sufficiently small to undergo Brownian motion. A recent computational study by Mari et al. \cite{mari2015discontinuous} suggests that features such as the importance of frictional particle contacts should carry over. This would satisfy a key requirement for observing shear jamming over a sizable range in packing fractions \cite{bi2011jamming}. Smith et al. \cite{smith2010dilatancy} indeed observed jamming in colloidal systems under extension where brittle fracture happens at the neck region rather than at the top plate. However, how jamming fronts propagate and a dynamic shear jammed state is established and maintained in the presence of Brownian motion remains to be explored.
\section{Acknowledgments}
We thank Eric Brown, Nicole James, Shomeek Mukhopadhyay, Qin Xu, Tom Witten and Wendy Zhang for stimulating discussions. This work was supported by the Chicago MRSEC, which is funded by NSF through grant DMR-1420709. Additional support was provided by ARO grant W911NF-16-1-0078. S.M. acknowledges support through a MRSEC Kadanoff-Rice fellowship.
\appendix
\section{Estimation of peak stress}

We estimate the average stress on the top plate due to solvent drainage through the fully saturated jammed granular bed from the Kozeny - Carman relationship.

The classic Kozeny - Carman (K-C) relationship is given by,
\begin{eqnarray}
\frac{\Delta P}{L} = \frac{180 \eta_s {\phi}^2 v_s}{{\Xi}^2 a^2 (1 - {\phi})^3}, 
\label{e5}
\end{eqnarray}
where $L$ is the thickness of the granular bed.
\newline
The drainage happens mainly through the paths of least resistance and for simplicity we assume that it happens along the top surface of the frustum in a region very close to the plate (see Fig. \ref{F14}). Under this situation, we need to take into account the variable path length $L$ as a function of $r$ and estimate the local pressure $P(r)$, which can then be integrated to estimate the average stress on the plate. 
\newline
The differential force on the top plate from the drainage (at a constant velocity $v_s$) through the ring of radius $r$ and width $dr$ (the distance of the ring circumference from the rim of the plate is $r_p - r$):
$dF(r) = P(r)\, 2 \pi r\,dr$ $\,\,$ which can be written as,
\begin{eqnarray}
dF(r) \approx \frac{180 \eta_s {\phi}^2 v_s (r_p - r)\,2 \pi r\,dr }{{\Xi}^2 a^2 (1 - {\phi})^3} = K\,(r_p - r) r\,dr, 
\end{eqnarray}

where $r_p$ is the plate radius and $K = \frac{180 \eta_s {\phi}^2 v_s\,2 \pi}{{\Xi}^2 a^2 (1 - {\phi})^3}$.
\newline
\newline
The total force on the top plate therefore is $F \approx \int_{0}^{r_p} K(r_p - r)\,r\,dr$
 
$= K\,[r_p\,{r}^2/2 - {r_p}^3/3]_{0}^{r_p} = K\,{r_p}^3 / 6$.
\newline
\newline
For the stress measured by pulling on the top plate this gives $\sigma \approx \frac{F}{\pi\,{r_p}^2} = K\,r_p / (6\,\pi)$.
\newline
Putting in the value of $K$ from Eq.(A2) we find $\sigma \approx \frac{60 \eta_s {\phi}^2 v_s\,r_p}{{\Xi}^2 a^2 (1 - {\phi})^3}$, which is the one third of the classic K-C value [Eq.(A1)] for the pressure drop over a granular bed of length $r_p$.
\newline
Putting in typical values we see that the drainage through the sidewall (i.e., through the path of least resistance) at a constant velocity corresponds to an average stress of $\approx$ 3 MPa in steady state. However, we note that along the rim [$L \sim a$ in Eq.(A1)] the plate can locally pull out some solvent at a much lower stress of $\approx$ 9 kPa. The breakage of the solvent at the rim can give rise to a dynamic delamination process as mentioned in the main text.

As we noted before, this scenario simplifies the actual situation and should only be taken as an estimate for the theoretical upper limit to the pulling force. Besides not accounting for the dynamical delamination process, the scenario also assumes that upper surface of the frustum is completely flat and in perfect contact with the plate, which may not always be the case.
This issue is underscored by the observation that the peak force depends to some degree on how far the pulling plate is pressed initially against the suspension surface.

However, the qualitative applicability of this scenario is confirmed by the observed scaling of the pulling force with pulling speed, packing fraction and solvent viscosity. This is shown in Fig. \ref{F11}.

\section{Calculation of peak stress from lubrication approximation}

The lubrication stress between two particles having a separation $H$ between surfaces of the particles is given by,
\begin{eqnarray}
\sigma = \frac{3\, \pi\, \eta_s\, v}{2 \,H}
\label{e3}
\end{eqnarray}
where $\eta_s$, the solvent viscosity = 8 mPa\,s and $v$ is the relative velocity between the particles.
\newline
\newline
We found that beyond a packing fraction $\phi \sim$ 0.5 for rice-starch particles, the suspension forms big clumps and becomes very difficult to mix. However, for cornstarch suspension mixing becomes difficult at a higher value of $\phi\, (\sim$ 0.56). Thus, we need to stay below this packing fractions to ensure the homogeneity of the suspension.
\newline
\newline
Although starch particles have irregular shapes, for the sake of calculation we assume them to be randomly distributed hard spheres having average diameter $a$. Under this assumption, for $\phi$ = 0.52, $H\sim 0.01\,a$ and for for $\phi$ = 0.46, $H\sim 0.02\,a$.
\newline
\newline
For cornstarch particles, a $\sim$ 12.5 $\mu$m and for rice-starch particles, a $\sim$ 2 $\mu$m. To estimate an upper limit of the stress predicted by lubrication theory, we assume $v$ to be the pulling velocity.
\newline
\newline
Putting in the values in Eq.(B1), we get the peak stress values for cornstarch suspension to be 2.3 kPa and 0.87 kPa for $v$ = 8 mm/s and 3 mm/s, respectively. For rice-starch, the peak stress for $v$ = 8 mm/s is 7.5 kPa and for $v$ = 3 mm/s, the peak stress is 2.8 kPa. These values are well below the experimentally observed peak stresses, as shown in Fig. \ref{F15}. These observations indicate that the lubrication forces between the particles are not setting the upper limits of the observed stresses under extension. 
\newline
\newline


\begin{thebibliography}{24}
\expandafter\ifx\csname natexlab\endcsname\relax\def\natexlab#1{#1}\fi
\expandafter\ifx\csname bibnamefont\endcsname\relax
  \def\bibnamefont#1{#1}\fi
\expandafter\ifx\csname bibfnamefont\endcsname\relax
  \def\bibfnamefont#1{#1}\fi
\expandafter\ifx\csname citenamefont\endcsname\relax
  \def\citenamefont#1{#1}\fi
\expandafter\ifx\csname url\endcsname\relax
  \def\url#1{\texttt{#1}}\fi
\expandafter\ifx\csname urlprefix\endcsname\relax\def\urlprefix{URL }\fi
\providecommand{\bibinfo}[2]{#2}
\providecommand{\eprint}[2][]{\url{#2}}

\bibitem[{\citenamefont{Waitukaitis and Jaeger}(2012)}]{waitukaitis2012impact}
\bibinfo{author}{\bibfnamefont{S.~R.} \bibnamefont{Waitukaitis}}
  \bibnamefont{and} \bibinfo{author}{\bibfnamefont{H.~M.}
  \bibnamefont{Jaeger}}, \bibinfo{journal}{Nature}
  \textbf{\bibinfo{volume}{487}}, \bibinfo{pages}{205} (\bibinfo{year}{2012}).

\bibitem[{\citenamefont{White et~al.}(2010)\citenamefont{White, Chellamuthu,
  and Rothstein}}]{white2010extensional}
\bibinfo{author}{\bibfnamefont{E.~E.~Bishoff} \bibnamefont{White}},
  \bibinfo{author}{\bibfnamefont{M.}~\bibnamefont{Chellamuthu}},
  \bibnamefont{and} \bibinfo{author}{\bibfnamefont{J.~P.}
  \bibnamefont{Rothstein}}, \bibinfo{journal}{Rheologica Acta}
  \textbf{\bibinfo{volume}{49}}, \bibinfo{pages}{119} (\bibinfo{year}{2010}).

\bibitem[{\citenamefont{Smith et~al.}(2010)\citenamefont{Smith, Besseling,
  Cates, and Bertola}}]{smith2010dilatancy}
\bibinfo{author}{\bibfnamefont{M.}~\bibnamefont{Smith}},
  \bibinfo{author}{\bibfnamefont{R.}~\bibnamefont{Besseling}},
  \bibinfo{author}{\bibfnamefont{M.}~\bibnamefont{Cates}}, \bibnamefont{and}
  \bibinfo{author}{\bibfnamefont{V.}~\bibnamefont{Bertola}},
  \bibinfo{journal}{Nature Communications} \textbf{\bibinfo{volume}{1}},
  \bibinfo{pages}{114} (\bibinfo{year}{2010}).

\bibitem[{\citenamefont{Smith}(2014)}]{smith2014fracture}
\bibinfo{author}{\bibfnamefont{M.}~\bibnamefont{Smith}},
  \bibinfo{journal}{Scientific Reports} \textbf{\bibinfo{volume}{5}},
  \bibinfo{pages}{14175} (\bibinfo{year}{2014}).

\bibitem[{\citenamefont{Peters and Jaeger}(2014)}]{peters2014quasi}
\bibinfo{author}{\bibfnamefont{I.~R.} \bibnamefont{Peters}} \bibnamefont{and}
  \bibinfo{author}{\bibfnamefont{H.~M.} \bibnamefont{Jaeger}},
  \bibinfo{journal}{Soft Matter} \textbf{\bibinfo{volume}{10}},
  \bibinfo{pages}{6564} (\bibinfo{year}{2014}).

\bibitem[{\citenamefont{Peters et~al.}(2016)\citenamefont{Peters, Majumdar, and
  Jaeger}}]{peters2016direct}
\bibinfo{author}{\bibfnamefont{I.~R.} \bibnamefont{Peters}},
  \bibinfo{author}{\bibfnamefont{S.}~\bibnamefont{Majumdar}}, \bibnamefont{and}
  \bibinfo{author}{\bibfnamefont{H.~M.} \bibnamefont{Jaeger}},
  \bibinfo{journal}{Nature} \textbf{\bibinfo{volume}{532}},
  \bibinfo{pages}{214} (\bibinfo{year}{2016}).

\bibitem[{\citenamefont{Cates et~al.}(1998)\citenamefont{Cates, Wittmer,
  Bouchaud, and Claudin}}]{cates1998jamming}
\bibinfo{author}{\bibfnamefont{M.}~\bibnamefont{Cates}},
  \bibinfo{author}{\bibfnamefont{J.}~\bibnamefont{Wittmer}},
  \bibinfo{author}{\bibfnamefont{J.-P.} \bibnamefont{Bouchaud}},
  \bibnamefont{and} \bibinfo{author}{\bibfnamefont{P.}~\bibnamefont{Claudin}},
  \bibinfo{journal}{Physical review letters} \textbf{\bibinfo{volume}{81}},
  \bibinfo{pages}{1841} (\bibinfo{year}{1998}).

\bibitem[{\citenamefont{Kumar and Luding}(2015)}]{kumar6167memory}
\bibinfo{author}{\bibfnamefont{N.}~\bibnamefont{Kumar}} \bibnamefont{and}
  \bibinfo{author}{\bibfnamefont{S.}~\bibnamefont{Luding}},
  \bibinfo{journal}{arXiv preprint arXiv:1407.6167}  (\bibinfo{year}{2015}).

\bibitem[{\citenamefont{Bi et~al.}(2011)\citenamefont{Bi, Zhang, Chakraborty,
  and Behringer}}]{bi2011jamming}
\bibinfo{author}{\bibfnamefont{D.}~\bibnamefont{Bi}},
  \bibinfo{author}{\bibfnamefont{J.}~\bibnamefont{Zhang}},
  \bibinfo{author}{\bibfnamefont{B.}~\bibnamefont{Chakraborty}},
  \bibnamefont{and}
  \bibinfo{author}{\bibfnamefont{R.}~\bibnamefont{Behringer}},
  \bibinfo{journal}{Nature} \textbf{\bibinfo{volume}{480}},
  \bibinfo{pages}{355} (\bibinfo{year}{2011}).

\bibitem[{\citenamefont{Vinutha and Sastry}(2016)}]{vinutha2016disentangling}
\bibinfo{author}{\bibfnamefont{H.}~\bibnamefont{Vinutha}} \bibnamefont{and}
  \bibinfo{author}{\bibfnamefont{S.}~\bibnamefont{Sastry}},
  \bibinfo{journal}{Nature Physics} \textbf{\bibinfo{volume}{12}},
  \bibinfo{pages}{578} (\bibinfo{year}{2016}).
	
	\bibitem[{\citenamefont{Seto et~al.}(2013)\citenamefont{Seto, Mari, Morris, and
  Denn}}]{seto2013discontinuous}
\bibinfo{author}{\bibfnamefont{R.}~\bibnamefont{Seto}},
  \bibinfo{author}{\bibfnamefont{R.}~\bibnamefont{Mari}},
  \bibinfo{author}{\bibfnamefont{J.~F.} \bibnamefont{Morris}},
  \bibnamefont{and} \bibinfo{author}{\bibfnamefont{M.~M.} \bibnamefont{Denn}},
  \bibinfo{journal}{Physical review letters} \textbf{\bibinfo{volume}{111}},
  \bibinfo{pages}{218301} (\bibinfo{year}{2013}).

\bibitem[{\citenamefont{Fall et~al.}(2015)\citenamefont{Fall, Bertrand,
  Hautemayou, Mezi{\`e}re, Moucheront, Lema{\^\i}tre, and
  Ovarlez}}]{fall2015macroscopic}
\bibinfo{author}{\bibfnamefont{A.}~\bibnamefont{Fall}},
  \bibinfo{author}{\bibfnamefont{F.}~\bibnamefont{Bertrand}},
  \bibinfo{author}{\bibfnamefont{D.}~\bibnamefont{Hautemayou}},
  \bibinfo{author}{\bibfnamefont{C.}~\bibnamefont{Mezi{\`e}re}},
  \bibinfo{author}{\bibfnamefont{P.}~\bibnamefont{Moucheront}},
  \bibinfo{author}{\bibfnamefont{A.}~\bibnamefont{Lema{\^\i}tre}},
  \bibnamefont{and} \bibinfo{author}{\bibfnamefont{G.}~\bibnamefont{Ovarlez}},
  \bibinfo{journal}{Physical Review Letters} \textbf{\bibinfo{volume}{114}},
  \bibinfo{pages}{098301} (\bibinfo{year}{2015}).

\bibitem[{\citenamefont{Wyart and Cates}(2014)}]{wyart2014discontinuous}
\bibinfo{author}{\bibfnamefont{M.}~\bibnamefont{Wyart}} \bibnamefont{and}
  \bibinfo{author}{\bibfnamefont{M.}~\bibnamefont{Cates}},
  \bibinfo{journal}{Physical Review Letters} \textbf{\bibinfo{volume}{112}},
  \bibinfo{pages}{098302} (\bibinfo{year}{2014}).

\bibitem[{\citenamefont{Brown and Jaeger}(2014)}]{brown2014shear}
\bibinfo{author}{\bibfnamefont{E.}~\bibnamefont{Brown}} \bibnamefont{and}
  \bibinfo{author}{\bibfnamefont{H.~M.} \bibnamefont{Jaeger}},
  \bibinfo{journal}{Reports on Progress in Physics}
  \textbf{\bibinfo{volume}{77}}, \bibinfo{pages}{046602}
  (\bibinfo{year}{2014}).

\bibitem[{\citenamefont{Brown and Jaeger}(2012)}]{brown2012role}
\bibinfo{author}{\bibfnamefont{E.}~\bibnamefont{Brown}} \bibnamefont{and}
  \bibinfo{author}{\bibfnamefont{H.~M.} \bibnamefont{Jaeger}},
  \bibinfo{journal}{Journal of Rheology} \textbf{\bibinfo{volume}{56}},
  \bibinfo{pages}{875} (\bibinfo{year}{2012}).

\bibitem[{\citenamefont{Fall et~al.}(2008)\citenamefont{Fall, Huang, Bertrand,
  Ovarlez, and Bonn}}]{fall2008shear}
\bibinfo{author}{\bibfnamefont{A.}~\bibnamefont{Fall}},
  \bibinfo{author}{\bibfnamefont{N.}~\bibnamefont{Huang}},
  \bibinfo{author}{\bibfnamefont{F.}~\bibnamefont{Bertrand}},
  \bibinfo{author}{\bibfnamefont{G.}~\bibnamefont{Ovarlez}}, \bibnamefont{and}
  \bibinfo{author}{\bibfnamefont{D.}~\bibnamefont{Bonn}},
  \bibinfo{journal}{Physical Review Letters} \textbf{\bibinfo{volume}{100}},
  \bibinfo{pages}{018301} (\bibinfo{year}{2008}).

\bibitem[{\citenamefont{Waitukaitis}(2014)}]{waitukaitis2014impact}
\bibinfo{author}{\bibfnamefont{S.~R.} \bibnamefont{Waitukaitis}},
  \emph{\bibinfo{title}{Impact-activated solidification of cornstarch and water
  suspensions}} (\bibinfo{publisher}{Springer}, \bibinfo{year}{2014}).

\bibitem[{\citenamefont{Han et~al.}(2016)\citenamefont{Han, Peters, and
  Jaeger}}]{han2016high}
\bibinfo{author}{\bibfnamefont{E.}~\bibnamefont{Han}},
  \bibinfo{author}{\bibfnamefont{I.~R.} \bibnamefont{Peters}},
  \bibnamefont{and} \bibinfo{author}{\bibfnamefont{H.~M.}
  \bibnamefont{Jaeger}}, \bibinfo{journal}{Nature Comm.} \textbf{\bibinfo{volume}{7}},
  \bibinfo{pages}{12243} (\bibinfo{year}{2016}).

\bibitem[{\citenamefont{Cates and Wyart}(2014)}]{cates2014granulation}
\bibinfo{author}{\bibfnamefont{M.~E.} \bibnamefont{Cates}} \bibnamefont{and}
  \bibinfo{author}{\bibfnamefont{M.}~\bibnamefont{Wyart}},
  \bibinfo{journal}{Rheologica Acta} \textbf{\bibinfo{volume}{53}},
  \bibinfo{pages}{755} (\bibinfo{year}{2014}).

\bibitem[{\citenamefont{Gadala-Maria and Acrivos}(1980)}]{gadala1980shear}
\bibinfo{author}{\bibfnamefont{F.}~\bibnamefont{Gadala-Maria}}
  \bibnamefont{and} \bibinfo{author}{\bibfnamefont{A.}~\bibnamefont{Acrivos}},
  \bibinfo{journal}{Journal of Rheology (1978-present)}
  \textbf{\bibinfo{volume}{24}}, \bibinfo{pages}{799} (\bibinfo{year}{1980}).

\bibitem[{\citenamefont{Lin et~al.}(2015)\citenamefont{Lin, Guy, Hermes, Ness,
  Sun, Poon, and Cohen}}]{lin2015hydrodynamic}
\bibinfo{author}{\bibfnamefont{N.~Y.} \bibnamefont{Lin}},
  \bibinfo{author}{\bibfnamefont{B.~M.} \bibnamefont{Guy}},
  \bibinfo{author}{\bibfnamefont{M.}~\bibnamefont{Hermes}},
  \bibinfo{author}{\bibfnamefont{C.}~\bibnamefont{Ness}},
  \bibinfo{author}{\bibfnamefont{J.}~\bibnamefont{Sun}},
  \bibinfo{author}{\bibfnamefont{W.~C.} \bibnamefont{Poon}}, \bibnamefont{and}
  \bibinfo{author}{\bibfnamefont{I.}~\bibnamefont{Cohen}},
  \bibinfo{journal}{Physical review letters} \textbf{\bibinfo{volume}{115}},
  \bibinfo{pages}{228304} (\bibinfo{year}{2015}).

\bibitem[{\citenamefont{McCabe et~al.}(2005)\citenamefont{McCabe, Smith, and
  Harriott}}]{mccabe2005adsorption}
\bibinfo{author}{\bibfnamefont{W.~L.} \bibnamefont{McCabe}},
  \bibinfo{author}{\bibfnamefont{J.~C.} \bibnamefont{Smith}}, \bibnamefont{and}
  \bibinfo{author}{\bibfnamefont{P.}~\bibnamefont{Harriott}},
  \emph{\bibinfo{title}{Unit operations of chemical engineering}}
  (\bibinfo{publisher}{McGraw-Hill, New York}, \bibinfo{year}{2005}).

\bibitem[{\citenamefont{von Kann et~al.}(2011)\citenamefont{von Kann, Snoeijer,
  Lohse, and van~der Meer}}]{von2011nonmonotonic}
\bibinfo{author}{\bibfnamefont{S.}~\bibnamefont{von Kann}},
  \bibinfo{author}{\bibfnamefont{J.~H.} \bibnamefont{Snoeijer}},
  \bibinfo{author}{\bibfnamefont{D.}~\bibnamefont{Lohse}}, \bibnamefont{and}
  \bibinfo{author}{\bibfnamefont{D.}~\bibnamefont{van~der Meer}},
  \bibinfo{journal}{Physical Review E} \textbf{\bibinfo{volume}{84}},
  \bibinfo{pages}{060401} (\bibinfo{year}{2011}).

\bibitem[{\citenamefont{Mari et~al.}(2015)\citenamefont{Mari, Seto, Morris, and
  Denn}}]{mari2015discontinuous}
\bibinfo{author}{\bibfnamefont{R.}~\bibnamefont{Mari}},
  \bibinfo{author}{\bibfnamefont{R.}~\bibnamefont{Seto}},
  \bibinfo{author}{\bibfnamefont{J.~F.} \bibnamefont{Morris}},
  \bibnamefont{and} \bibinfo{author}{\bibfnamefont{M.~M.} \bibnamefont{Denn}},
  \bibinfo{journal}{Proceedings of the National Academy of Sciences}
  \textbf{\bibinfo{volume}{112}}, \bibinfo{pages}{15326}
  (\bibinfo{year}{2015}).

\end{thebibliography}
\end{document}